\documentclass[a4paper,12pt]{article}
\usepackage{amsmath,graphicx,subfigure}       

\def\e{\epsilon}
\def\g{\gamma}
\def\m{\mu}
\def\n{\nu}
\def\r{\rho}
\def\s{\sigma}
\def\no{\nonumber}
\def\oh{\frac{1}{2}}
\def\pa{\partial}

\begin{document}
\begin{titlepage}
\begin{flushright}
IFUM 652/FT \\
December 1999 \\
\end{flushright}
\vspace{1.5cm}
\begin{center}
{\bf \large SEMI-NAIVE DIMENSIONAL RENORMALIZATION }
\footnote{Work supported in part by M.U.R.S.T.}\\
\vspace{1 cm}
{ M. PERNICI} \\ 
\vspace{2mm}
{\em INFN, Sezione di Milano, Via Celoria 16, I-20133 Milano, Italy}\\
\vspace{1cm}
\bf{ABSTRACT}
\end{center}
\begin{quote}

~~ We propose a treatment of $\g^5$ in dimensional regularization
which is based on an algebraically consistent extension of the 
Breitenlohner-Maison-'t Hooft-Veltman (BMHV) scheme;
 we define the corresponding minimal renormalization scheme
and show its equivalence with a non-minimal BMHV scheme.

The restoration of the chiral Ward identities requires the introduction of 
considerably fewer finite counterterms than in the BMHV  scheme.
This scheme is the same as the minimal
naive dimensional renormalization 
in the case of diagrams not involving fermionic traces with an odd number
of $\g^5$, but unlike the latter it is a consistent scheme. 

As a simple example we apply our minimal subtraction scheme to the
Yukawa model at two loops in presence of external gauge fields.

\end{quote}
\end{titlepage}
\section*{Introduction}

Dimensional regularization \cite{HV,DR}
is the most powerful scheme for making 
perturbative computations in quantum field theory. It is particularly
effective in absence of chiral symmetries, since minimal subtraction
\cite{tH} preserves the vectorial gauge symmetries.

In \cite{HV} it has been shown that defining $\g^5$ as in four
dimensions the axial anomaly is reproduced.
In \cite{BM} it has been proven that this definition of $\g^5$
leads to a consistent regularization scheme (BMHV)
at all orders in perturbation theory.

The restoration of the chiral Ward identities in the BMHV scheme has
been studied in \cite{Bonn1}.

In the BMHV scheme many of the computational advantages of the dimensional
regularization scheme are lost, since 
the restoration of the chiral
Ward identities requires the introduction of non-invariant
counterterms for practically any diagram containing $\g^5$.

Due to this difficulty, comparatively 
few quantities  have been  computed in this scheme. 
Among them, there are some one-loop processes  in the standard
model \cite{Korner}, two-loop anomalous dimensions  of
four-fermion operators in QCD \cite{BW}, the three-loop
anomalous dimension of axial currents in QCD \cite{Larin}.
The Adler-Bardeen non-renormalization theorem \cite{AB}
has been verified at two loops in this scheme in \cite{Bos}.

Systematic computations of the bare action in the BMHV scheme
have been made in non-abelian gauge theories with
chiral fermions \cite{Martin}, where the finite one-loop counterterms
necessary to restore the Slavnov-Taylor identities have been computed;
in the simplest Yukawa model at two loop in the minimal scheme \cite{Sch};
in a general class of Yukawa models at two loops, where 
 the finite two-loop counterterms necessary to restore the rigid Ward 
identities have been computed, together with some of those 
necessary to recover the local Ward identities in presence of external
gauge fields \cite{PRR}.

Most of the computations in chiral theories are made in some
version of the naive dimensional regularization scheme (NDR),
first considered in \cite{BGL} and \cite{CFH}; other proposals
of NDR have been made in \cite{GD,AZ}. 
The common feature of all these versions of NDR is that in 
fermionic loops with  an even number of $\g^5$'s,  these are
considered to be anticommuting  with all the gamma matrices,
thus avoiding  the spurious anomalies present in the BMHV scheme.
However all these regularization
schemes have some algebraic inconsistency.
To avoid these algebraic inconsistencies
it has been suggested that the cyclic property of the trace be
upheld \cite{Town}, but  in this case bosonic symmetry  is not
respected, and no consistency proof for these renormalization
schemes is available.

Reviews on various proposals for dimensional
regularization in chiral theories  can be found in \cite{Bonn,Baikov} .

In this paper we introduce a consistent regularization scheme,
in which to  the usual $\gamma_{\mu}$ and $\g^5$ formal objects of the
BMHV scheme two new $\g^5$-like objects $\eta_1$  and $\eta_2$ are
added; the former is not anticommuting with all the gamma matrices,
as $\g^5$, while the latter is anticommuting;
 they are defined in such a way that traces with an even 
number of $\eta_1$ or an odd number of $\eta_2$ vanish.

We will call this scheme semi-naive dimensional regularization (SNDR);
due to its algebraic consistency in $d=4-\e$ dimension, algebraic
manipulations on the bare quantities can be done unambiguously.

We define minimal  subtraction (MS) in this scheme in such a way that,
after taking the limit $\e \to 0$ on the minimally
subtracted Green functions, there exists a homomorphism 
from these abstract subtracted Green functions to the usual 
renormalized Green functions, which preserves the algebraic properties
on convergent diagrams, like the skeleton expansion.

We prove the equivalence of the MS-SNDR scheme with a non-minimal BMHV
scheme.




The proof of the consistency of the MS-SNDR scheme is the main result
of this paper. We do not deal with
the practical problem of restoring the
Ward identities, apart from an explicit example and 
some general considerations, showing that in this scheme there are 
considerably fewer spurious anomalies than in the MS-BMHV scheme.

As an example of the practical
convenience of our scheme with respect to the BMHV scheme, we treat in
some
detail the two-loop renormalization of the Yukawa model in presence of
external gauge currents, which we have worked out in the BMHV scheme in
\cite{PRR}; while in the latter scheme minimal subtraction does not
preserve the (anomalous) chiral Ward identities, 
so that finite counterterms must be added to restore them for every
relevant term,
in the SNDR scheme, after performing minimal subtraction  only 
two finite two-loop counterterms are needed to
 preserve the anomalous
Ward identities, with the anomaly appearing in the Bardeen form \cite{Bard};
the Adler-Bardeen theorem \cite{AB} is verified.

We discuss briefly the case of chiral gauge theories and we
compare SNDR  with previously proposed NDR schemes.

\section{Semi-naive dimensional regularization and minimal subtraction}

\subsection{Breitenlohner-Maison-'t Hooft-Veltman scheme}

We recall how gamma matrices are treated in the 't Hooft and Veltman 
\cite{HV} dimensional regularization scheme as elaborated by 
Breitenlohner and Maison (BMHV) \cite{BM} .
We use Euclidean space conventions.

In the BMHV scheme one considers the Lorentz covariants  $I$, 
$\delta_{\mu \nu}$, $\g_\m$, $p_\m$, etc. as formal objects, satisfying the
usual tensorial rules.   
 $\delta_{\mu \nu}$ is the Kronecker delta in
$d=4-\epsilon$ dimensions; a formal rule for summed indices is given;
\begin{equation}
\delta_{\mu \nu} \delta_{\nu \r} = \delta_{\m\r} ~~;~~
\delta_{\mu \nu} p_{\nu} = p_{\mu}~~;~~\delta_{\mu \mu} = d 
\end{equation}
The Lorentz indices of these formal covariants do not take a specific
value.

The gamma `matrices' $\gamma_{\mu}$  satisfy the
relation
\begin{equation}
\{ \gamma_{\mu},\gamma_{\nu} \} = -2 \delta_{\mu \nu} I
\end{equation}
where $I$ is the identity,
\begin{equation}
I^2 = I ~~;~~ I \gamma_{\mu} = \gamma_{\mu} I = \gamma_{\mu}
\end{equation}
The trace is cyclic and satisfies
\begin{equation}\label{trace}
tr~ I = 4 
\end{equation}

In the BMHV scheme additional `$(d-4)$-dimensional' or `evanescent'
tensors $\hat{\delta}_{\mu \nu},\hat{p}_{\mu}$ and $\hat{\gamma}_{\mu}$ 
are introduced; the Kronecker delta in the $(d-4)$-dimensional space
is $\hat{\delta}_{\mu \nu}$ , satisfying 
\begin{eqnarray}
&&\hat{\delta}_{\mu \nu}  \delta_{\nu \rho} = 
\hat{\delta}_{\mu \nu}  \hat{\delta}_{\nu \rho} = 
\hat{\delta}_{\mu \rho}~~;~~
\hat{\delta}_{\mu \mu} = - \epsilon \nonumber \\
&&\hat{p}_{\mu} \equiv \hat{\delta}_{\mu \nu} p_{\nu} ~~;~~
\hat{\gamma}_{\mu} \equiv \hat{\delta}_{\mu \nu} \gamma_{\nu} 
\end{eqnarray}
The Kronecker delta in four dimensions in $\bar{\delta}_{\m\n}$,
satisfying
\begin{eqnarray}
\bar{\delta}_{\m\n} \equiv \delta_{\m\n} - \hat{\delta}_{\mu \nu}
 ~~;~~
\bar{p}_\m  \equiv \bar{\delta}_{\m\n} p_\n  ~~;~~
\bar{\g}_\m  \equiv \bar{\delta}_{\m\n} \g_\n
\end{eqnarray}

The Levi-Civita antisymmetric tensor has no evanescent component:
\begin{equation}
\hat{\delta}_{\mu \nu} \epsilon_{\nu \rho \sigma \tau} = 0 
\end{equation}
and satisfies
\begin{equation}
\epsilon_{\m_1 \m_2 \m_3 \m_4} 
\epsilon_{\n_1 \n_2 \n_3 \n_4} 
= \sum_{\pi  \in S_4} sign (\pi) \Pi_{i=1}^4 
\bar{\delta}_{\m_i \n_{\pi (i)}}
 \end{equation}
The `matrix' $\g^5$ is defined by
\begin{equation}\label{g5}
\g^5 = \frac{1}{  4!} \epsilon_{\mu \nu \rho \sigma} \gamma_{\mu}
\gamma_{\nu} \gamma_{\rho} \gamma_{\sigma}
\end{equation}
which implies
\begin{eqnarray}
&&\{ \g^5, \gamma_{\mu} \} = 2 \hat{\gamma}_{\mu} \g^5 \\
&&\g^5 \g^5 = I~~;~~ [\g^5,\hat{\gamma}_{\mu}] = 0
\end{eqnarray}

After having performed the Dirac algebra with these rules, 
reduced the number of Levi-Civita tensors,
eliminated traces, reduced products of gamma matrices to the
corresponding antisymmetric products and eliminated summed 
indices, the BMHV elements are reduced to normal form.
The Feynman graphs are expressed in terms of meromorphic functions
of $\e$, which are the coefficients of these normal form terms.
 
 The minimal subtraction on a $l$-loop $1PI$ Feynman graph renormalized
up to order $l-1$ consists in subtracting local terms, 
polynomial in momenta and masses, which are the poles of the
Laurent series in $\e$ of the
meromorphic functions defining such a graph, both for hatted and
non-hatted tensor structures in normal form.

The bare action is a formal object, from which the Feynman graphs are 
constructed following the usual rules. 
In the minimal subtraction scheme the
bare action $l$-loop counterterms are the poles of the 
$l$-loop $1PI$ Green functions,
computed using the Feynman rules derived from the $(l-1)$-loop bare action.

After making the minimal subtraction, 
the limit for $\e \to 0$ is taken in the entire functions which are
the coefficients of the normal form terms,
and all the hatted tensors are set to zero; finally the
formal sums, traces and the 
normal form covariants are identified with the 
usual four-dimensional operations and objects respectively.

In Euclidean space the reflection symmetry takes the place of
hermiticity \cite{Collins}.
 Reflection symmetry is an antilinear involution, with
\begin{eqnarray}
&&\Theta \psi (x) = \bar{\psi}(x') \gamma_1 ~~;~~
\Theta \bar{\psi}(x) = \gamma_1 \psi (x') \\
&&\Theta  \g_\m = - \g_\m ~~~~;~~~~ \Theta  \g^5 = \g^5 
\end{eqnarray}
where $x^{'1}=-x^1$, $x^{'\mu} = x^{\mu}$ for $\mu \neq 1$.
$\delta_{\m\n}$, $\e_{\m\n\r\s}$ and $\e$ are invariant under
reflection symmetry.

Considering  a multiplet of fermions,
the following fermionic bilinears are reflection symmetric: 
\begin{eqnarray}\label{dec}
&&\int \bar{\psi}(H_1 \gamma_{\mu} + H_2 \hat{\gamma}_{\mu} +
H_3 \g^5 \bar{\g}_{\mu}  +
 i H_4 \g^5 \hat{\gamma}_{\mu} ) \partial_{\mu} \psi  +
\bar{\psi} (i H_5 A +  H_6 \g^5 B) \psi  \no
 \\
&&+ \bar{\psi} (i H_7  \gamma_{\mu} V_{\mu} +
i H_8  \g^5 \bar{\g}_{\mu}  A_{\mu} +
i H_9 \hat{\gamma}_{\mu} V_{\mu} +
H_{10} \g^5 \hat{\gamma}_{\mu} A_{\mu}) \psi
\end{eqnarray}
where the scalar $A$, the pseudoscalar $B$, the vector $V_{\mu}$ and
the pseudovector $A_{\mu}$ are real fields, with
\begin{eqnarray}
\Theta A(x) = A(x') ~;~~ 
\Theta B(x) = B(x') ~;~~ 
\Theta V_{\mu}(x) = V_{\mu}'(x')  ~;~~ 
\Theta A_{\mu}(x) = A_{\mu}'(x') \no
\end{eqnarray}
and $H_i$ are hermitian
matrices commuting with the gamma matrices.

A non-minimal subtraction consists in performing, 
after the minimal subtraction, the subtraction of  finite local
counterterms; to avoid ambiguities, one must characterize which are
the normal form terms on which the finite subtractions are
made.

In the case of a reflection symmetric theory, a convenient choice of
fermionic bilinears finite counterterms are the non-hatted terms in
(\ref{dec}). The advantage of this choice is that in the case of
non-chiral theories the $d$-dimensional Lorentz symmetry of the bare
action is preserved.

\newpage

\subsection{Extension of the BMHV algebra}

Add to the BMHV Dirac algebra the objects $\eta$ and $\eta_1$
 satisfying the following defining relations, 
assumed to be valid for arbitrary $\e$:
\begin{eqnarray}
\label{A1}&&\{ \eta, \gamma_{\mu} \} = 2 \hat{\gamma}_{\mu} \eta_1 \\
\label{A2}&&\eta^2 = I \\
\label{A21}&&I \eta = \eta \\
\label{A3}&&[ \eta_1, \hat{\gamma}_{\mu} ] = 0  \\
\label{A31}&&[ \eta_1, \eta ] = 0  \\
\label{A4}&&tr~ \eta \gamma_{\mu} \g_\n \g_\r \g_\s = 
4 ~\epsilon_{\mu \nu \rho \sigma} \\
\label{A5}&&tr~ \eta_1^2 = 0 
\end{eqnarray}
We assume that
the trace has  the property (\ref{trace})  and that it is
cyclic on this enlarged algebra.

Under reflection symmetry, $\Theta \eta = \eta$ and 
$\Theta \eta_1 = \eta_1$.

Define
\begin{eqnarray}
\label{eta2}\eta_2 \equiv \eta - \eta_1
\end{eqnarray}
Consider first the case $\e \neq 0$.

Let us prove that
\begin{eqnarray}
\label{B0}&&\eta I = \eta \\
\label{B1}&&\{ \eta_1, \gamma_{\mu} \} = 2 \hat{\gamma}_{\mu} \eta_1 \\
\label{B2}&&\{ \eta_2, \gamma_{\mu} \} = 0 \\
\label{B21}&& \eta_1 I =  I \eta_1 = \eta_1 \\
\label{B3}&&\eta_1 \eta_2 = \eta_2 \eta_1 = 0 \\
\label{B4}&&\eta_1^2 + \eta_2^2 = I
~~;~~\eta_1^3 = \eta_1 ~~;~~\eta_2^3 = \eta_2 
\end{eqnarray}
Eq. (\ref{B0}) is obvious.
From (\ref{A1})  one gets
\begin{eqnarray}
\label{B5}&&\e \eta_1 = \frac{1}{2} \hat{\g}_\m \{ \eta, \g_\m \} =
\oh \e \eta + \frac{1}{2} \hat{\g}_\m \eta \hat{\g}_\m 
\end{eqnarray}
Using (\ref{A1}), (\ref{A3}), (\ref{eta2}) and (\ref{B0})  one gets 
 the relations (\ref{B1}), (\ref{B2}) and (\ref{B21}).
From (\ref{A31}), (\ref{B1}) and (\ref{B2}) one gets easily
\begin{eqnarray}
&&[\eta_1^2,\g_\m]= [\eta_2^2,\g_\m]= 0 \no \\
\label{B7}&&\g_\m \eta_1 \eta_2 \g_\m = (d-8) \eta_1 \eta_2  \\
&&[\eta_1,\eta_2] = 0 \no
\end{eqnarray}
From (\ref{A2}), (\ref{eta2}) and (\ref{B7}) one has
\begin{eqnarray}
&&-d = \g_\m \eta^2 \g_\m =
\g_\m (\eta_1^2 + \eta_2^2 + \{ \eta_1,\eta_2 \} ) \g_\m = \no \\
&&-d (\eta_1^2 + \eta_2^2) + (d-8) \{ \eta_1,\eta_2 \} \no
\end{eqnarray}
so that
\begin{eqnarray}
\label{B8}\e \{ \eta_1,\eta_2 \} =0
\end{eqnarray}
Since $\e \neq 0$, we get $\{ \eta_1,\eta_2 \} = 0$, which together with
(\ref{A31}) implies (\ref{B3}).

Finally (\ref{B4}), which follows trivially from  (\ref{A2}) 
and (\ref{B3}), says that 
$\eta_1^2$ and $\eta_2^2$ are orthogonal projectors commuting with all
the elements of this algebra.

Let us prove the following relations for the trace:
\begin{eqnarray}
\label{B9}&&tr~ \eta_1 = tr~ \eta_2 = 0 \\
\label{B10}&&tr~ \eta_1^2 \g_{\m_1} ... \g_{\m_r} = 0 \\
&&tr~\eta_2^2 \g_{\m_1} ... \g_{\m_r} =  tr~ \g_{\m_1} ... \g_{\m_r} \\
\label{B12}&&tr~ \eta_2 \g_{\m_1} ... \g_{\m_r} = 0 \\
\label{B11}&&tr~\eta_1 \g^5 \g_{\m_1} ... \g_{\m_r} = 
 tr~ \g_{\m_1} ... \g_{\m_r} 
\end{eqnarray}
for arbitrary choice of indices $\m_1,..,\m_r$, $r \geq 1$.
One has
\begin{eqnarray}
&&-d ~tr~\eta_1 = tr~ \eta_1 \g_\m  \g_\m =
tr~ \g_\m \eta_1 \g_\m = (8-d) tr~\eta_1 \no \\
&&-d ~tr~\eta_2 = tr~ \eta_2 \g_\m  \g_\m =
tr~ \g_\m \eta_2 \g_\m = d ~tr~\eta_2 
\end{eqnarray}
from which (\ref{B9}) follows.
To prove the remaining relations (\ref{B10}-\ref{B11}), observe that
if for all $\m$
\begin{eqnarray}
X_{\pm} \g_\m = \pm \g_\m X_{\pm}
\end{eqnarray}
then, for indices $\m_1,..,\m_r$ all different
\begin{eqnarray}\label{dproof}\label{dpr}
&&-d~tr ~X_{\pm} \g_{\m_1} ... \g_{\m_r} =
tr ~X_{\pm} \g_{\m_1} ... \g_{\m_r} \g_\m  \g_\m = \no \\
&&\pm tr ~X_{\pm} \g_\m \g_{\m_1} ... \g_{\m_r} \g_\m =
\pm (-1)^r (2r-d) tr ~X_{\pm} \g_{\m_1} ... \g_{\m_r}
\end{eqnarray}
which for non-integer $d$ has solution only for
\begin{eqnarray}
tr ~X_{\pm} \g_{\m_1} ... \g_{\m_r} =0
\end{eqnarray}
Using this observation and the fact that $\eta_1^2$, $\eta_1 \g^5$ and
$\eta_2^2$ commute with all the gamma matrices, while $\eta_2$
anticommutes with all of them,
one gets relations (\ref{B10}-\ref{B11})
for indices $\m_1,..,\m_r$ all different.
In particular one gets 
\begin{eqnarray}\label{B125}
tr~\eta_2 \g^5 = 0
\end{eqnarray}

The general case for (\ref{B10}-\ref{B11})
follows from this particular case since, simplifying
gamma matrices with equal indices, one is reduced to the relations
\begin{eqnarray}
tr~ \eta_1^2 = tr~ \eta_2 = 0 ~~~;~~tr~ \eta_2^2 =4 ~~~;~~ tr~\eta_1 \g^5 = 4
\end{eqnarray}
which follow trivially from the relations
(\ref{A2}-\ref{A5}) and from (\ref{B9}), (\ref{B125}).

Using (\ref{B9}-\ref{B11}) all the traces can be computed, observing
that the traces with more than two $\eta_1$ or two $\eta_2$ can be
reduced to the cases (\ref{B9}-\ref{B11}) using (\ref{A1}-\ref{B4}).

Observe that in $d \neq 4$ dimensions this algebra of covariants does not
include $\e_{\m_1...\m_d}$, as might be thought due to the presence of
$\eta_2$, which is a $\g^5$-like object anticommuting with all gamma
matrices. In fact
$tr~\eta_2 \g_{\m_1} ... \g_{\m_d} = 0$ due to (\ref{B12}).

In the case $\e = 0$,  $\hat{\g}_\m$ vanishes and  
$\eta_1$  is an independent generator, since
eq.(\ref{B5}) becomes trivial; in
that case, add the defining relations
\begin{eqnarray}
&&\{ \eta_1, \bar{\gamma}_{\mu} \} = 0 \\
&&\eta \eta_1 = \eta_1^2 \\
&&I \eta_1 = \eta_1 I = \eta_1
\end{eqnarray}
which we obtained from (\ref{A1}-\ref{A31}) in the case $\e \neq 0$.

\vskip 1cm

In the case $\e = 0$, after identifying the formal sums 
(e.g. $p_\m \g_\m$) as usual Einstein convention sums,
 there is a homomorphism $\phi$ between the subalgebra 
generated by $\g_\m = \bar{\g}_\m$ and $\eta$, 
with the trace $tr$, and
the usual four-dimensional Dirac algebra generated by 
$\g_\m'$, with unit matrix $I'$ and  with the usual trace $tr'$:
This homomorphism is uniquely determined by mapping the generators,
\begin{eqnarray}
\phi (\g_\m) = \g_\m' ~~;~~\phi (\eta) = 
\g^{5'} \equiv \frac{1}{4!} \e_{\m\n\r\s} \g_\m' \g_\n' \g_\r' \g_\s'
\end{eqnarray}
and by the relations defining the homomorphism
\begin{eqnarray}\label{ho}
&&\phi (I) = I' \no \\
&&\phi (X_1 X_2) = \phi (X_1) \phi (X_2) ~~;~~
 \phi (a X_1 + b X_2) = a \phi (X_1) + b \phi (X_2) \no \\
&&tr~ X = tr'~ \phi (X)
\end{eqnarray}
for $X, X_1$ and $X_2$ belonging to this subalgebra; $a$ and $b$ are
complex numbers.
$p_\m$, $\delta_{\m\n}$ and $\e_{\m\n\r\s}$ are 
mapped into the corresponding usual tensors in four dimensions.

To verify that this is a homomorphism with the properties (\ref{ho})
it is sufficient to verify them on the  relations 
(\ref{trace},\ref{A1},\ref{A2},\ref{A21},\ref{A4}) defining the
subalgebra for $\e = 0$.

The kernel of this homomorphism is obtained
 projecting  this subalgebra with $(I -\eta \g^5)/2$, which commutes
 with all the elements of the subalgebra.

$\g_\m (I +\eta \g^5)/2$ generates the orthogonal subalgebra,
establishing an isomorphism between this subalgebra and the usual
Dirac algebra.

This homomorphism cannot be extended to $\eta_1$, since
\begin{eqnarray}
4 = tr~ (\g^5 - \eta) \eta_1 \neq tr'~ (\g^{5'}-\g^{5'}) \phi (\eta_1)= 0 \no
\end{eqnarray}
for any value of $\phi (\eta_1)$.

Furthermore it cannot be extended to the case $\e \neq 0$; for
instance 
\begin{eqnarray}\label{exw}
8 \e ~\bar{\delta}_{\m\n} = 
tr (\g^5 - \eta)\bar{\g}_\m  \g_\r \eta \g_\r \g_\n \neq
tr'~ (\g^{5'}-\g^{5'})\phi (\bar{\g}_\m \g_\r \eta  \g_\r \g_\n) = 0 
\end{eqnarray}
for any value of $\phi (\g_\m)$.  

\vskip 1cm

In order to check the consistency of the defining relations
(\ref{A1}-\ref{A5}),
let us give an explicit representation of this extension of the Dirac
algebra for $d = 2 n$ even dimensions.
Denote by $\g_\m'$ the usual $2^{d/2}$-dimensional gamma matrices;
define further $\g^{5'} \equiv \g_1' \g_2' \g_3' \g_4'$ and
$\eta' \equiv i^{(d-1)d/2} \g_1' ... \g_d'$ 
(the $\g_5'$ in the right-hand-side of the latter expression 
is $\g_\m$ for $\m = 5$ and should not
be confused with the pseudoscalar object called $\g^{5'}$ defined
in the former expression).

The extended algebra elements are given by the following tensor
products:
\begin{eqnarray}
&&I \equiv I' \times diag (1,1,1,1) \no \\
&&\g_\m \equiv \g_\m' \times diag (1,1,1,1) \no  \\
&&\g^5 \equiv \g^{5'} \times diag (1,1,1,1) \\
&&\eta_1 \equiv \g^{5'}  \times diag (1,-1,0,0) \no  \\
&&\eta_2 \equiv \eta'  \times diag (0,0,1,-1)  \no \\
&&tr~ X = 2^{1- \frac{d}{2}} Sp [ (I' \times diag (1,-1,1,1)) X ]  \no 
\end{eqnarray}
where $Sp$ is the ordinary  trace on $2^{d/2 + 2} \times 2^{d/2 + 2}$
matrices,
while $tr$ is a not a matrix trace, but it is a cyclic multilinear
operator on the space generated by the gamma matrices, $\eta$ and
$\eta_1$.

In this representation all the defining relations for the extended
Dirac algebra hold. All the properties which we derived  above hold in
this representation, 
 including (\ref{B12}), which however does not
follow  from (\ref{dpr})  in the case $r= 2 n$.

\subsection{Semi-naive dimensional regularization}

In the semi-naive dimensional regularization (SNDR)
the Feynman amplitudes are expressed in terms of meromorphic functions
of the complex parameter $d=4-\e$; in the case of fermionic
amplitudes, there is a meromorphic function for each element
of the generalized gamma
matrix algebra described above, written in normal form (NF). 

A basis for the gamma matrix normal form quantities belonging to the
generalized Dirac algebra consists in 
\begin{eqnarray}
&&\{ \Gamma_A, \eta \Gamma_A, \eta_1 \Gamma_A, 
\eta \eta_1  \Gamma_A\} \\
&&\{  \Gamma_A \} = \{ I, \g_\m, 
\g_{[\m_1} \g_{\m_2]},.., \g_{[\m_1}... \g_{\m_r]},...\}
\end{eqnarray}
where $[...]$ indicates antisymmetrization.

It can be decomposed in the following way:
\begin{eqnarray}
&&NF = NF_0 + NF_1 + \hat{NF} \\
&&NF_0 =  \{\bar{\Gamma}_A, \eta \bar{\Gamma}_A \} \\
&&NF_1 =  \{\eta_1 \bar{\Gamma}_A, \eta \eta_1 \bar{\Gamma}_A \} \\
&&\{ \bar{\Gamma}_A \} =  \{ I, \bar{\g}_\m, 
\bar{\g}_{[\m} \bar{\g}_{\n]}, \g^5 \bar{\g}_\m, \g^5 \}
\end{eqnarray}
$\hat{NF}$ consists of the remaining
NF terms, which have at least a $\hat{\g}_\m$.

Denote by $A_0$ and $A_1$ the subalgebras generated by 
$NF_0$ and $NF_1$ respectively, on the complex numbers (constant in $\e$).

Denote by 
$\hat{A}$  the subalgebra generated by $\hat{NF}$,  multiplied
by an entire function in $\e$, and by any NF term multiplied by an entire
function with a zero in $\e = 0$.
The objects in $\hat{A}$ are evanescent.

One has 
\begin{eqnarray}
&&A_0~A_0 = A_0 ~~;~~A_0 ~A_1 = A_1 A_0 = A_1  ~A_1 = A_1 \no
\\
&&\hat{A} ~A_0 = A_0 \hat{A} = 
\hat{A} ~A_1 = A_1 \hat{A} = \hat{A}
\end{eqnarray}

Denote by $T_0$ the algebra on constants in $\e$, whose 
generators are the normal form terms in $NF_0$, the momenta
$\bar{p}_\m$, $\delta_{\m\n}$, $\e_{\m\n\r\s}$
and the normal forms obtained
from these taking tensor products of these (e.g. $\bar{\delta}_{\m\n}$,
$I,~\bar{p}_\m I \times I, \bar{\delta}_{\m\n} I \times I$,
$\bar{\g}_\m \times \eta$
are normal forms in $T_0$; these tensor products are introduced to
cover the case of Feynman graphs with more than one open fermionic line).

Denote by $T_1$ the algebra on constants in $\e$, whose 
generators are the normal form terms in $NF_1$ and the normal
forms obtained taking the tensor product of at least one $NF_1$ 
term and $T_0$ normal forms.

Finally denote by $\hat{T}$ the space formed by normal forms with an
hatted quantity, on constants in $\e$, and by any normal form 
multiplied by an entire function with a zero in $\e = 0$,
and by the elements obtained making a tensor product of at least one
of these elements and those previously defined in $T_0$ and $T_1$.
 This is the algebra of evanescent quantities.
One has 
\begin{eqnarray}\label{T1}
&&T_0~T_0 = T_0 ~~;~~T_0 ~T_1 = T_1 T_0 = T_1  ~T_1 = T_1 \no
\\
&&\hat{T} ~T_0 = T_0 \hat{T} = 
\hat{T} ~T_1 = T_1 \hat{T} = \hat{T}
\end{eqnarray}
The homomorphism $\phi$ defined in the previous subsection
extends canonically to $T_0$ , but cannot be extended to
$T_1$ or to $\hat{T}$.

After the subtraction procedure, 
a renormalized but still regularized $1PI$ Feynman graph
$G_n^{\e ren}$, written in normal form, belongs to the
abstract algebra $T_0+T_1 +\hat{T}$.  
The subscript $n$ indicates the Lorentz and internal indices of the
Feynman graph.

The algebraic consistence of the extended BMHV algebra guarantees that
it is equivalent to compute $G_n^{\e ren}$ directly in terms of
convergent integrals, or to compute separately the divergent integrals for
 $G_n^{\e unren}$ and 
its counterdiagram graphs, and then sum them; the latter procedure is
much easier.

$G_n^{\e ren}$
must be evaluated as a four-dimensional quantity $G_n^{ren}$,
expressed in terms of the usual Dirac algebra.
 This is done in two steps:

i) set $\e$ and the hatted objects to zero; denote by 
$G_n^{\e = 0 ren}$ the resulting quantity, belonging to
$T_0+T_1$;

ii) Map the abstract objects in the generalized
BMHV algebra to the usual Dirac algebra quantities:
\begin{eqnarray}
\phi (G_n^{\e = 0 ren}) = G_n^{ren}
\end{eqnarray}
where $G_n^{ren}$ is the renormalized $1PI$ graph after the
regulator has been removed.
The map $\phi$ must be the trace-preserving homomorphism previously defined.

It is necessary that this map be a trace-preserving homomorphism
to preserve the
algebraic properties on convergent diagrams, like 
making a skeleton expansion of a convergent $1PI$ Green function; 
these expansions involve making products of relevant $1PI$ vertices and dressed
propagators, sums over Lorentz indices and Dirac traces, which are
preserved by a homomorphism preserving the traces.
It is clear that this homomorphism is unique.
Since it cannot be defined on $T_1$, it follows that 
 a consistent renormalization scheme in the SNDR scheme must
be such that {\it the} $T_1$ {\it terms are absent from } $G_n^{\e = 0 ren}$
at all orders in loops (and hence from the $\e^0$ term of the Laurent
series for $G_n^{\e  ren}$, which therefore must belong to
$T_0 +\hat{T}$, while $G_n^{\e = 0 ren}$ must belong to $T_0$).

This is a non-trivial requirement, since the $T_1$-dependent counterterms
can only be local, while non-local terms are produced in perturbation
theory; for instance in a unsubtracted two-loop fermionic
two-point function (without its 
one-loop counterdiagrams) there are generally 
both non-local poles ( this is true also in
the BMHV case) and non-local finite $A_1$ terms; the point is that
these non-local terms are canceled by the one-loop counterdiagrams.
In the next subsection we will show that this non-trivial property
holds at all orders in perturbation theory.

The second step is trivial in the BMHV scheme, since the BMHV Dirac
algebra for $\e = 0$ is isomorphic with the usual Dirac algebra in
four dimensions.
This is not so in the SNDR Dirac algebra for $\e = 0$.

We will call a quantity, belonging to  the extended  Dirac  algebra,
 to be {\it weakly evanescent}  if it belongs to the kernel of the
homomorphism $\phi$, modulo evanescent terms. The space of
weakly evanescent quantities is called $W$.

The kernel of $\phi$ on the domain $A_0$ is $\frac{1-\eta \g^5}{2} A_0$.
The subspace $\frac{1+\eta \g^5}{2} A_0$ is isomorphic to the usual
Dirac algebra. Clearly
\begin{eqnarray}\label{weak}
W A_0 = W W = W
\end{eqnarray}

In the general tensor product case
\begin{eqnarray}\label{weak2}
W T_0 = W W = W
\end{eqnarray}

Summarizing:
the regularized quantities (bare action, $G_n^{\e ren}$)
  live in a space with four times as many Dirac
algebra elements as in BMHV;
$G_n^{\e = 0 ren}$ lives in a space with twice as many Dirac
algebra elements as in BMHV;
$G_n^{\e ren}$ differs from $G_n^{\e = 0 ren}$ by
evanescent quantities;
$G_n^{ren}$ is isomorphic to $G_n^{\e ren}$ modulo weakly
evanescent quantities; projecting each $A_0$ component 
of $G_n^{\e = 0 ren}$ to $\frac{1+\eta \g^5}{2} A_0$
one gets an object isomorphic to  $G_n^{ren}$.

\subsection{Minimal subtraction in SNDR}

The renormalization by minimal subtraction  (MS) consists in a minimal
subtraction procedure and in the  evaluation in four dimensions of  the
formal objects defined in dimensional regularization.

We define it recursively starting from $l=1$ loops.

Denote by $G_n^{\e (l)}(k)$ a  $l$-loop
$1PI$ Feynman graph   with momenta $k$, minimally
renormalized according to our MS rules up to
$l' < l$ loops but yet unrenormalized at $l$ loops.
Denote by $\Gamma_n^{\e (l)}(k)$ the corresponding Green function,
which is the sum of various $G_n^{\e (l)}(k)$.

We  define the minimal subtraction procedure for the SNDR in the following way.

The minimal subtraction on $G_n^{\e (l)}(k)$
 consists in subtracting all the local terms, 
polynomial in momenta and masses, which are:

i) the poles of the Laurent expansion of the
meromorphic functions defining such a graph, both for hatted and
non-hatted tensor structures in normal form;

ii) the $T_1$ terms

The resulting renormalized but yet regularized $1PI$ Feynman graph
$G_n^{\e (l) ren}$ is evaluated in four dimensions following the
procedure described in the previous subsection.

The $l$-loop bare action counterterms consist in the local terms
subtracted from the $1PI$ $l$-loop Green functions 
$\Gamma_n^{\e (l)}(k)$
according to these rules.

In a renormalizable theory without insertions of
composite operators of dimension
larger than five, the only $T_1$ counterterms are  $A_1$
terms (there are no relevant four-fermion terms).
 
As in MS-BMHV, we have to show that $G_n^{\e (l) ren}$ thus
defined
has no non-local poles; furthermore we have to show that it has no
non-local $T_1$ term.

Consider $G_n^{'\e (l)}(k)$ and
$\Gamma_n^{'\e (l)}(k)$ to be the corresponding quantity in the
BMHV scheme.

 Suppose by induction hypothesis that in the BMHV scheme
finite local counterterms can be chosen for the renormalized 
(but still regularized) relevant $1PI$ Green functions
$\Gamma_{n'}^{'\e ren (l')}$ up to $l'=l-1$ loops such that
 $\Gamma_{n'}^{\e ren  (l')}-\Gamma_{n'}^{'\e  ren (l')}$ 
is weakly evanescent, so that
this difference vanishes after evaluation in four dimensions (in general the 
MS-SNDR scheme is equivalent to a non-minimal BMHV scheme).
 
Let us show that the same can be done at $l$ loops.

Let $d_n$ be the superficial degree of divergence of $G_n^{\e (l)}(k)$.
For $r > d_n$, $\partial_k^r G_n^{\e (l)}(k)$ and
$\partial_k^r G_n^{'\e (l)}(k)$
admit a skeleton expansion.

$\partial_k^r$ denotes $r$ derivatives with respect to the external
momenta $k$, with indices not saturated neither among themselves nor
with the Lorentz indices included in the subscript $n$.

The difference between the renormalized and still regularized
Green functions in the two schemes at $l' < l$ loops is weakly
evanescent by induction hypothesis, so that using the skeleton
expansion property and (\ref{T1},\ref{weak2})
it follows that for $r > d_n$ also
\begin{eqnarray}\label{diff}
\partial_k^r G_n^{\e (l)}(k) - \partial_k^r G_n^{'\e (l)}(k)
\in W
\end{eqnarray}
This equation  implies that
$G_n^{\e (l)}(k) - G_n^{'\e (l)}(k)$
is a polynomial of degree $d_n$ in  the momenta
plus a weakly evanescent quantity.
The polynomial part depends on the poles in $\e$ (of degree less or
equal to $l$) and on the normal forms, including the $T_1$ terms.
An important point in this kind of arguments \cite{CK}
is that $\partial_k^r$
commutes with the operations of extracting the poles or the $\e^0$
terms of the Laurent series on which the meromorphic functions are
expanded, and that it commutes also with the operation of projecting the $T_0$
elements to the kernel of $\phi$ and with the extraction of the
$T_1$ terms.

Making the minimal subtraction on $G_n^{\e (l)}(k)$ using our rules
and making an appropriate non-minimal subtraction in the BMHV sense on 
$G_n^{'\e (l)}(k)$, the polynomial part of 
$G_n^{\e (l)}(k) - G_n^{'\e (l)}(k)$ can be subtracted and the
 difference between the renormalized
$1PI$ Feynman graphs, yet regularized, is weakly evanescent.

We have therefore proven by induction that the MS-SNDR scheme is
equivalent to a (generally non-minimal) BMHV scheme 
at all orders in perturbation theory.

The proof of polynomiality in masses of the minimal subtraction counterterms 
proceeds in the same way \cite{Coll} and holds also for the 
$T_1$ counterterms.

Let us show from another point of view 
that step (ii) is necessary to define our minimal subtraction
scheme; consider a different definition of minimal subtraction for
SNDR, in which only poles are subtracted, and in which the
 evaluation procedure in four dimensions  includes the rule of setting
$\eta_1$  to zero and identifying $\eta$ with $\g^5$
(this map is not a trace-preserving homomorphism, as we saw in 
Subsection $1.2$).
Let us repeat the above argument, in which now $G_n^{\e (l)}$ refers
to this modified scheme, while $G_n^{'\e (l)}$ has the same
interpretation as above.
In this case the renormalization parts $\Gamma_{n'}^{\e ren(l')}$ of
the skeleton expansion of 
$\partial_k^r G_n^{\e (l)}$ differ from the corresponding renormalization
parts
in the BMHV scheme by $T_1$ terms and by weakly
evanescent terms. 
Performing a trace involving a weakly evanescent renormalization part 
and a $T_1$ contribution in another 
renormalization part of the skeleton expansion,
one gets in general a finite non-local $T_0$
contribution, which is absent in
$\partial_k^r G_n^{'\e (l)}(k)$.
In such a case it is not possible to choose $l$-loop local counterterms
to get the same renormalized Green functions in the two schemes,
up to weakly evanescent terms or local $T_1$ terms. 
Therefore one cannot go from the new scheme to the BMHV scheme with
a renormalization group transformation. This is not possible, so that
the proposed scheme is wrong.

The necessity of the  subtraction of the $T_1$ terms  to  have
a  consistent renormalization scheme makes our subtraction scheme
minimal; the polynomiality in the masses of these 
$T_1$ terms  supports the notion that MS-SNDR
is really an MS scheme.

A characteristics of the usual MS scheme is that to compute the 
$l$-loop counterterms one has to compute 
the $O(1/\e)$ part of $l$-loop Feynman integrals, the $O(1)$
 part of $(l-1)$-loop Feynman integrals, the $O(\e)$
 part of $(l-2)$-loop Feynman integrals, and so on.
In MS-SNDR the same happens.
In fact the $T_1$ $l$-loop terms in $\Gamma_n^{\e (l)}(k)$ arise
from a counterterm graph with a  counterterm
containing a $NF_1$ factor ($T_1$ terms are
absent at tree level), from a graph with  vertices containing
only $NF_0$ factors or from a graph containing at least a
vertex with a $\hat{NF}$ factor.

In the first case the Feynman integrals are clearly of $(l-1)$-loop order;
in the second case a $NF_1$ term is produced, in the reduction of the
graph to normal form, by anticommuting $\eta$
with $\g_\m$, generating a $\hat{\g}_\m \eta_1$ factor; to get a $T_1$
term this $\hat{\g}_\m$ must be contracted with another $\hat{\g}_\m$,
producing a $\e$, so that only the pole part of the $l$-loop Feynman
integral is involved in producing $T_1$; the same is true in the third
case.

Actually the subtraction of the $l$-loop $T_1$ terms in 
$\Gamma_n^{\e (l)}(k)$
can be done algebraically, without doing any analytic computation; in
fact one can subtract algebraically all the $NF_1$ terms (both poles
and finite parts) from $\Gamma_n^{\e (l)}(k)$, subtracting
subsequently the remaining poles. 
(In practice this algebraic subtraction can be done simply by
treating $\eta$ in
the open fermionic lines as anticommuting with all the gamma matrices
and by setting $\eta_1$ to zero on the same lines).
The analytic computation of the $T_1$ $l$-loop counterterms matters
only for computing  $\Gamma_n^{\e (l+1) ren}(k)$.

On the contrary, the subtraction of the $l$-loop poles from
$\Gamma_n^{\e (l)}(k)$ requires some analytic computation, so that
from the analytic point of view the $T_1$ counterterms require 
little effort with respect to the pole counterterms.

Let us discuss briefly finite renormalization, which are needed to recover the
chiral symmetry or more in general to renormalize the Green functions
in a non-minimal way, e.g. at a momentum subtraction point.

They belong to $T_0$, with the possible addition of $\hat{T}$ terms to
have  counterterms respecting $d$-dimensional Lorentz symmetry in the
non-chiral cases; the discussion is analogous to the one made in 
the first subsection for the non-minimal BMHV scheme.

If the tree-level action is reflection symmetric, the bare action in 
MS-SNDR is also reflection symmetric. 

In a non-minimal SNDR scheme, reflection symmetry is preserved by
choosing reflection symmetric $T_0$ finite renormalization terms.

Obviously non-minimal $T_1$  counterterms are forbidden.

\newpage

\section{Applications}
After having shown that the MS-SNDR is a consistent renormalization
scheme, let us show that it is convenient from a computational point
of view.

In this scheme the renormalized Green functions are the same as in
MS-NDR as long as fermionic loops with an odd number of $\g^5$ does
not occur. In these cases the computations are exactly the same in
the two schemes and the chiral Ward identities are satisfied.

In the remaining cases, in MS-SNDR  the traces with
odd number of $\eta$ or $\eta_1$ enter in the game ; in
these cases one must check explicitly the chiral Ward identities;
 if they are broken, they can be recovered as usual by
adding local finite
 counterterms to the MS counterterms, order by order in perturbation
 theory. These are the cases in which NDR becomes ambiguous or inconsistent.

In the case of external vector gauge fields,
the vectorial Ward identities are preserved in the MS-SNDR as well as
in the MS-BMHV scheme; in fact, since the vectorial gauge
transformations do not
involve neither the dimension, nor $\eta$ or $\eta_1$, the operations
of extracting the poles and the $T_1$ terms commute with gauge
transformations, so that the bare action is vector gauge invariant.
The extension of this argument to gauge theories will be discussed later.

\subsection{Yukawa models }

In the Yukawa models it is possible to use NDR with anticommuting
$\g^5$
without finding inconsistencies; MS-NDR preserves the rigid chiral
symmetries.
However coupling these  models to external gauge fields and considering
the corresponding Ward identities, NDR is not  anymore a consistent
regularization scheme, due to the presence of the chiral anomalies.

In the MS-BMHV scheme the chiral Ward identities are broken by a large
number of spurious  terms, which must be subtracted by introducing
non-invariant local finite counterterms in the bare action.
This has been done at one-loop and partly at two loops  in \cite{PRR},
choosing renormalization conditions which, in the cases in which
fermionic loops with an odd number of $\g^5$ do not occur, give the
same renormalized Green functions as in MS-NDR. 

In SNDR the tree-level bare action is
\begin{eqnarray}
&&S^{(0)} = \int \oh (D_\m \phi_i)^2  +
\oh \bar{\psi} \g_\m D_\m \psi -
\oh (D_\m \bar{\psi}) \g_\m \psi + \\
&&i \bar{\psi} y_i \psi \phi_i +
\frac{1}{4!} h_{ijkl} \phi_i \phi_j \phi_k \phi_l
\end{eqnarray}
We have set the dimensional regularization scale $\mu$ to one.  

We use the same notation as in \cite{PRR}, with the difference that in
the tree-level action and in the gauge transformations on the fermions
we replace $\g^5$ with $\eta$; the difference between the two
cases being weakly evanescent, this is an equivalent starting point
for making perturbation theory. We restrict our attention to the
massless Yukawa theories.
Furthermore we do not require that $\hat{A}_\m^a = 0$, which  was
chosen in \cite{PRR} to reduce the number of counterterms in the 
BMHV bare action.

 The fermions have internal indices which are not indicated
($\psi_I$).  The matrices $S_i,P_i$  are hermitian. 

We will consider a group $G$ which is not necessarily semi-simple.

The fermions  belong to a (reducible) representation $t^a$;
under an infinitesimal transformation
\begin{eqnarray} 
\delta \psi = i \e^a t^a \psi
\end{eqnarray}
We will consider chiral representations
\begin{eqnarray}\label{tap}
&&t^a = t_R^a P_R + t_L^a P_L = t_s^a I + t_p^a \eta
~~;~~ 
\bar{t}^a = t_L^a P_R + t_R^a P_L =
 t_s^a I - t_p^a \eta \no \\
&&t_R^a = t_s^a + t_p^a ~~;~~t_L^a = t_s^a  - t_p^a  \no \\
&&P_R = \frac{I+\eta}{2} ~~;~~
P_L = \frac{I-\eta}{2}
\end{eqnarray} 
where $t_R^a$ and $t_L^a$ belong in general to different representations. 

The Dirac conjugate fermion transforms as
\begin{eqnarray}
\delta \bar{\psi} = -i \e^a \bar{\psi} \bar{t}^a
\end{eqnarray}

The scalars  belong to a real (reducible) representation $(\theta^a)_{ij}$;
under an infinitesimal transformation 
\begin{eqnarray}
\delta \phi_i = i \e^a \theta^a_{ij} \phi_j
\end{eqnarray} 
The covariant derivatives are
\begin{eqnarray}
&&D_\m \psi = (I \pa_\m + i A_\mu^a t^a) \psi \no \\
&&D_\m \bar{\psi}  = \pa_\m \bar{\psi} - 
i A_\mu^a \bar{\psi} \bar{t}^a  \no \\
&&D_\m \phi_i = (\pa_\m \delta_{ij} + 
i A_\mu^a \theta^a_{ij}) \phi_j 
\end{eqnarray}

The Yukawa coupling $\bar{\psi} y^i \psi \phi_i$, with
\begin{eqnarray}
&&y_i = S_i I + i P_i \eta = Y^i P_R + Y^{i \dagger} P_L \no \\
&&Y_i = S_i + i P_i
\end{eqnarray}
 is invariant under these transformations provided
\begin{eqnarray}\label{yinv}
&&y^j \theta^a_{ji} + y^i t^a - \bar{t}^a y^i = 0 
\end{eqnarray}
or equivalently
\begin{eqnarray}
&&Y^j \theta^a_{ji} + Y^i t_R^a - t_L^a Y^i = 0 ~~;~~
Y^{j \dagger} \theta^a_{ji} + Y^{i \dagger} t_L^a - t_R^a Y^{i \dagger} = 0
\no
\end{eqnarray}

The tree-level action is reflection symmetric.

We use a modified subtraction scheme \cite{BL} , 
in which the bare constants are chosen of the form 
\begin{eqnarray}\label{ct}
&&c_A =  \sum_{l \geq 1} \hbar^l N_d^l
c^{(l)}_A(\epsilon) \nonumber \\
&& N_d = (4 \pi)^{\epsilon/2-2} \Gamma (1+\frac{\epsilon}{2})~~;~~
c^{(l)}_A(\epsilon) = \sum_{r \geq 0} c^{(l)}_{A,-r} \epsilon^{-r}
\end{eqnarray} 
For all counterterms which do not involve $NF_1$, in minimal
subtraction $c^{(l)}_A(\epsilon)$ has no $\e^0$ term.

The considerations made in the first section are easily generalized to
the modified subtraction schemes.

Let us discuss  the MS-SNDR renormalization at one and two loops,
comparing it with the MS-NDR computations in the MS-NDR scheme
\cite{Mac} and in the non-minimal BMHV scheme \cite{PRR}.

Consider first the scalar Green functions; since it is not possible to
produce renormalization parts involving $\e_{\m\n\r\s}$, these Green
functions have only fermionic loops with an even number of $\eta$'s,
which behave in these traces as
the anticommuting objects $\eta_2$. Therefore
in this case our MS scheme is the  same as the MS-NDR scheme; only
poles are subtracted, respecting the chiral Ward identities.
All the spurious terms present in BMHV \cite{PRR} are absent.

The same is true for the $\e_{\m\n\r\s}$-independent parts of the
Green functions involving external gauge fields with or without 
external scalar fields.
There is none of the spurious terms present in BMHV, with or without
$\hat{A}_\m^a$ (the former can be avoided in the case of
external gauge fields setting
$\hat{A}_\m^a = 0$, as in \cite{PRR}, but they must be included when
the gauge fields are promoted to quantum fields,
leading to many other spurious terms in BMHV, besides those 
computed in \cite{PRR}).
Let us now consider the relevant fermionic Green functions.
At one loop, multiplying them by $\eta_2^2$, they become identical to
those in NDR (after the replacement $\g^5 \to \eta_2$),
 and are minimal subtracted as in that case;
multiplying them by $\eta_1^2$, they become identical to those in
BMHV (after the replacement $\g^5 \to \eta_1$). 

Expressing the sum of the two sectors in the base of NF terms
involving $\eta$ and $\eta_1$, MS consists in subtracting, besides the
poles, the finite $NF_1$ terms, that is the $A_1$ terms.
Projecting the MS subtracted graphs with $\eta_2^2$, one has 
MS-NDR; projecting the MS subtracted graphs with $\eta_1^2$,
one has BMHV graphs renormalized in a non-minimal way (in the BMHV
sense), with a finite subtraction such that in four dimensions
the renormalized graphs are the same as in MS-NDR. This is precisely
the renormalization choice made in \cite{PRR}.

To renormalize the fermionic Green functions at two loops in our scheme
it is sufficient to multiply them by $\eta_2^2$ and perform MS-NDR.
The terms $S^{(2)}_{NF_1}$ are useful only to compute 
renormalized Green functions at three or more loops, so we will not
compute them here; let us only remark that all contributions 
to $S^{(2)}_{NF_1}$ apart
from those  with a fermionic loop subdiagram can be read out from 
the corresponding spurious terms in \cite{PRR}, 
replacing $\g^5$ with $\eta_1$ and projecting them with $\eta_1^2$.
The difference in the remaining  diagrams is that in  \cite{PRR}
the fermionic loop contains $\g^5$, not $\eta_2$, leading to some
difference in the coefficients of those counterterms, but to the same
renormalized Green functions.

The last renormalization parts to be considered are those involving
$\e_{\m\n\r\s}$, which are those with fermionic loops involving an odd
number of $\eta$'s, which behave as the non-naive $\g^5$-objects $\eta_1$.
At one loop the relevant graphs are the 
$\e_{\m\n\r\s}$-dependent parts of the three-vector and four-vector
graphs, which are the same as in BMHV, since $\eta_1$ acts as $\g^5$
in these cases. In our MS scheme, there are no spurious terms in the
bare action, corresponding to the fact that the anomaly appears in the
Bardeen form \cite{Bard}, preserving vectorial gauge invariance
(for a discussion, see \cite{PRR}).

At two loops, the $\e_{\m\n\r\s}$-dependent part of the three-vector
graphs involves again $\eta_1$; the fermionic counterterms are
therefore those obtained projecting the one-loop fermionic
counterterms
with $\eta_1^2$, that is are the same as in BMHV in   \cite{PRR}.
Therefore the two-loop anomaly terms cancel as in \cite{PRR},
in agreement with the Adler-Bardeen theorem \cite{AB}.

The complete bare action in MS-SNDR scheme is
\begin{eqnarray} 
&&S = \int \oh c_{ij} D_\m \phi_i D_\m \phi_j + 
\oh \bar{\psi} \g_\m (P_R c_{\psi} + P_L \bar{c}_{\psi})  D_\m \psi - 
\no \\
&&\oh (D_\m \bar{\psi}) (P_L c_{\psi} + P_R \bar{c}_{\psi}) \g_\m \psi
+ i \bar{\psi} c_i \psi \phi_i + \no \\
&&\frac{1}{4!} c_{ijkl} \phi_i \phi_j \phi_k \phi_l +
\frac{1}{4} c_{ab} F_{\m\n}^a F_{\m\n}^b + S_{NF_1} + \Delta S[A_\m^a]
\end{eqnarray} 
where
\begin{eqnarray}
&&\bar{c}_{\psi} \equiv \bar{c}_{\psi}(Y_i,Y_i^{\dagger}) 
\equiv c_{\psi}(Y_i^{\dagger},Y_i)
\no \\
&&c_{\psi} \equiv c_{\psi}(Y_i,Y_i^{\dagger})
\end{eqnarray} 
are hermitian due to reflection symmetry
and $\Delta S[A_\m^a]$ is the non-naive part of the pure gauge part of the
bare action. It is due to Feynman graphs containing one or more
 fermionic traces with odd number of $\eta$ or $\eta_1$, giving each a
Levi-Civita tensor.


At one loop one has
\begin{eqnarray}
&&c_{ij}^{(1)} = -\frac{4}{\e} Y_{ij} \no \\
&&c_{\psi}^{(1)} = -\frac{1}{\e} Y_i^{\dagger}  Y_i  \no \\
&&c_i^{(1)} = \frac{2}{\e} y_j y_i^{\dagger} y_j \no \\
&&c_{ijkl}^{(1)} = -\frac{48}{\e}  Y_{(ijkl)}
+ \frac{3}{\e}  h_{(ij}^{rs} h^{rs}_{kl)} \no \\
&&c_{ab}^{(1)} =  - \frac{8}{3 \e} S_2(F)^{ab} - \frac{1}{3 \e} S_2(S)^{ab}
\end{eqnarray}
and 
\begin{eqnarray}\label{fineta1}
&&S_{NF_1}^{(1)} = \int 
-\frac{i}{\e}\bar{\psi} \eta_1 \hat{\g}_\m P_i y_i D_\m \psi +
\frac{i}{\e}(D_\m \bar{\psi}) \eta_1 y_i P_i \hat{\g}_\m
\psi + \no \\
&&\bar{\psi} \eta_1 y_j P_i y_j \psi \phi_i -
i\bar{\psi} \eta_1 y_i t_p^a y_i^{\dagger} \bar{\g}_\m \psi A_\m^a \no \\
&&\Delta S^{(1)}[A_\m^a] = 0
\end{eqnarray}
where  $Y_{i_1i_2...i_{2n-1}i_{2n}}$, $S_2(F)$ and $S_2(S)$ are
defined as in \cite{PRR}, as well as the group covariant
$K_2^{ab}$  in the next formula.

We checked explicitly that
$S_{NF_1}$ is invariant under vectorial gauge transformations, i.e.
under transformations satisfying the constraint $\e^a t_p^a = 0$
(see (\ref{tap})),
as expected by the argument given at the beginning of this section.

At two loops one has \cite{Mac,PRR}
\begin{eqnarray}
&&c_{ij}^{(2)} =  
-\frac{1}{12\e}  h_{ikmn}h_{jkmn} +
(-\frac{8}{\e^2} + \frac{2}{\e}) Y_{ikjk} + 
(-\frac{4}{\e^2} + \frac{3}{\e}) Y_{ijkk}
\no \\
&&c_{\psi}^{(2)} =  
(-\frac{1}{2\e^2} + \frac{1}{8\e})
Y_j^{\dagger} Y_i Y_i^{\dagger} Y_j -
\frac{2}{\e^2} Y_j^{\dagger} Y_i Y_j^{\dagger} Y_i+ 
(-\frac{2}{\e^2} + \frac{3}{2\e})
 Y_{ij}~Y_j^{\dagger}  Y_i 
\no \\
&&c_i^{(2)} =  
(\frac{4}{\e^2} - \frac{2}{\e}) Y_{jk} y_k y_i^{\dagger} y_j+ 
(\frac{2}{\e^2} - \frac{1}{\e}) y_k y_j^{\dagger} y_i y_j^{\dagger}y_k
+
\nonumber \\
&&(\frac{1}{\e^2} - \frac{1}{2\e})
y_k (y_i^{\dagger} y_j y_j^{\dagger} +y_j^{\dagger} y_j y_i^{\dagger}) y_k+
\frac{2}{\e^2}
y_k (y_i^{\dagger} y_j y_k^{\dagger} + y_j^{\dagger} y_k y_i^{\dagger})y_j+
\nonumber \\
&& \frac{1}{\e} y_k y_j^{\dagger} y_i y_k^{\dagger}y_j -
  \frac{1}{\e} h_{ijkl}y_k  y_j^{\dagger} y_l
\no \\
&&c_{ijkl}^{(2)} = 
(-\frac{192}{\epsilon^2}+\frac{96}{\epsilon}) Y_{ninjkl} + 
\frac{48}{\epsilon} Y_{nijnkl}+
(-\frac{96}{\epsilon^2}+\frac{48}{\epsilon}) Y_{nijkln} + \nonumber \\
&&h_{mnij}[
-\frac{96}{\epsilon^2}  Y_{mkln} + 
(-\frac{48}{\epsilon^2}+\frac{24}{\epsilon}) Y_{mknl} +
(\frac{12}{\epsilon^2}-\frac{6}{\epsilon}) Y_{mp} h_{pnkl} + \nonumber \\
&&\frac{3}{\epsilon^2} h_{mnpq} h_{pqkl} + 
(\frac{6}{\epsilon^2} - \frac{3}{\epsilon}) h_{npqk} h_{mpql}] \no \\
&&c_{ab}^{(2)} = -\frac{2}{\e} K_2^{ab}
\end{eqnarray}
where in the expression for $c_{ijkl}^{(2)}$ the symmetrization in the
indices $i,j,k,l$ is understood.

We didn't compute $\Delta S^{(2)}[A_\m^a]$, which might be different
from zero; in that case, it must be an $\e_{\m\n\r\s}$ term, and it
is the same as in BMHV scheme used in \cite{PRR}.
Observe that  $\Delta S^{(2)}[A_\m^a]$ does not have to do with the anomaly,
since it does not involve the totally symmetric tensor $d^{abc}$.

We didn't compute $S_{NF_1}^{(2)}$ which, as explained in the previous
section, is not necessary to compute the two-loop renormalized
Green functions.

In absence of the external gauge fields MS-SNDR preserves the rigid
chiral Ward identities at all loops.
 In fact it is not possible to produce the relevant
terms with  $\e_{\m\n\r\s}$, which are at the origin of the
breaking of the symmetries. 

In presence of the external gauge fields the chiral Ward identities
might  be broken by the
$\e_{\m\n\r\s}$ contributions, so that a finite counterterm 
$\Delta S^{(2)}[A_\m^a]^{non~min}$ might be necessary to recover the 
chiral Ward identities; at more than two loops
$\Delta S^{(l)}[A_\m^a]$ can contain also parity even terms, coming
from an even number of Levi-Civita tensors.

\vskip 1cm

Let us finally discuss what would have happened if, as discussed 
at the end of the
previous section, we had not subtracted the $A_1$ terms.
\begin{figure}
\centering
\includegraphics[width=3cm]{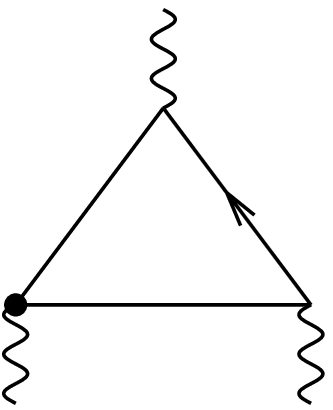}
\caption{}
\label{fig:1}
\end{figure} 

In the computation of the
anomaly, the difference from the correctly MS-subtracted case and the
one now discussed consists of the counterterm graph in Figure
\ref{fig:1}, in which there is the insertion of the $A_1$ counterterm
in (\ref{fineta1}), 
leading to a non-local finite contribution, which conspires to the
absence of two-loop corrections to the anomaly; apart from non satisfying
the Adler-Bardeen theorem, the suggested
theory would have the following inconsistency:
in the one-loop approximation
it has the same renormalized Green functions as in a consistent
renormalization scheme, the non-minimal BMHV scheme used in \cite{PRR}; 
from general renormalization theory
arguments it follows that the renormalized Green functions in
two different renormalization schemes can be connected by a renormalization
group transformation; being the one-loop relevant terms equal, it
follows that the marginal Green functions can differ at two loops only by
a local term; however we just found a case in which this difference is 
non-local; this is not possible, so that the proposed scheme is
inconsistent,
in agreement with the arguments in the previous subsection.

\subsection{Gauge theories}

In the Yukawa theories only the pure gauge Green functions do not
satisfy the chiral Ward identities in the MS-SNDR scheme;
the remaining Green functions satisfy them at all orders in
perturbation theory. The vector Ward identities are respected in this
scheme.

In chiral gauge theories no Green function satisfies the chiral Ward
identity in this scheme (at order high enough in loops); however
SNDR breaks the chiral symmetries in a gentler way than BMHV.

\begin{figure}
\centering
\subfigure[]{\label{fig:2.1}\includegraphics[width=2.5cm]{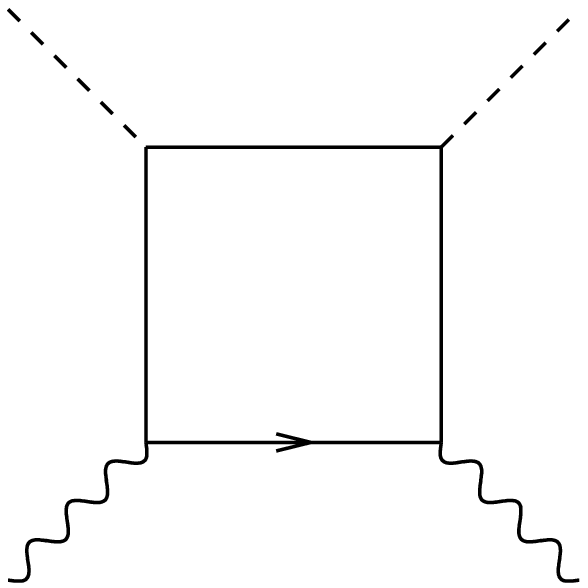}} 
\hspace{1cm}
\subfigure[]{\label{fig:2.2}\includegraphics[width=3cm]{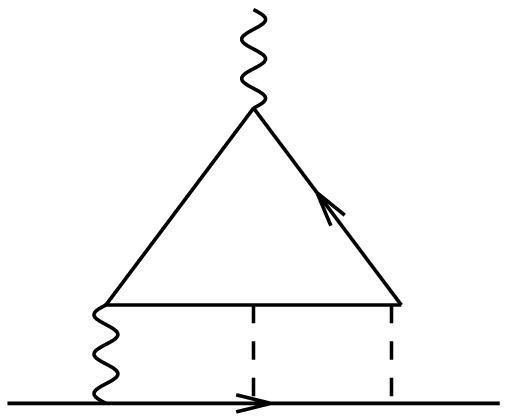}}
\hspace{1cm}
\subfigure[]{\label{fig:2.3}\includegraphics[width=3cm]{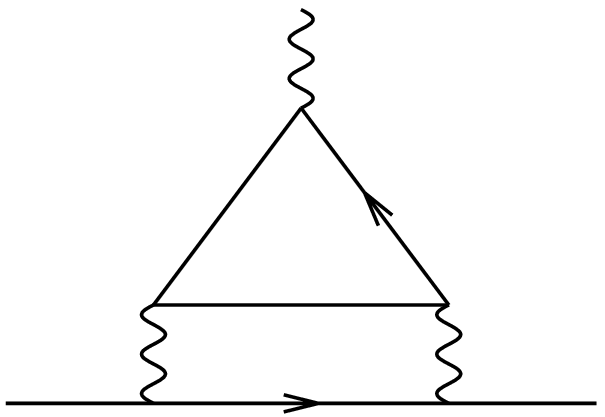}}
\caption{}
\label{fig:2}
\end{figure}

 As long as graphs involving fermionic loops with an
odd number of $\g^5$-like objects are absent, the chiral Ward
identities will be satisfied. In general, we can expect that the level
at which MS is not sufficient in SNDR is the same at which NDR reveals
its inconsistencies. Apart from the graphs mentioned in the previous section,
 it is reasonable to expect that other
superficially divergent graphs
containing the subgraph in Figure \ref{fig:2.1} (giving non-vanishing
contributions to the two-loop anomaly \cite{PRR}),
like the one  in Figure \ref{fig:2.2}  need to be subtracted in a
non-minimal way.

In presence of composite operators the need of introducing a
finite renormalization appears already at low loop order; for instance
coupling an external gauge current to the theory, the 
minimally subtracted triangle graph
involving it and two chiral gauge fields gives the anomaly in the Bardeen
form which does not respect the chiral
Ward identities  so that a finite counterterm must be added to restore
them. 

Apart from this issue,
the MS-SNDR scheme for a chiral gauge theory differs qualitatively from
that in Yukawa theories for the appearance of $\g^5$ itself (not $\eta$ or 
$\eta_1$) in the bare action.
It is only at this point that the bare action starts living in the
extended algebra four times bigger than the BMHV algebra, as can be
seen observing that, for instance, in absence of counterterms with $\g^5$
 in a chiral gauge theory the relevant bare 
vectorial part of the vertices 
contains $\g_\m$ and  $\eta \eta_1 \g_\m$,  but not $\eta \g^5 \g_\m$ and
$\eta_1 \g^5 \g_\m$.

For example consider the graph in Figure \ref{fig:2.3}, 
contributing to the vertex
for an external gauge field (it is chosen to be external to avoid
anomaly cancellation); in MS it generates counterterms involving
$\frac{1}{3!}\e_{\m\n\r\s} \g_\n \g_\r \g_\s/\e = \g^5 \bar{\g}_\m/\e$, 
while at lower loop level only vertices involving $\eta$ and $\eta_1$
appeared in the bare action. Similar counterterms can be expected in
the graph in Figure \ref{fig:2.2}; this counterterm cannot be replaced
by $\eta \bar{\g}_\m/\e$, since the difference is not weakly
evanescent (see the example in eq. (\ref{exw})).

As far as vector gauge invariance is concerned, let us argue that
MS-SNDR preserves it in the case of a non-abelian vector gauge theory
with chiral Yukawa couplings (i.e. the gauged version of the Yukawa
model in the previous subsection, with $t_p^a=0$).

The bare action is BRST-invariant at tree level. Suppose that it is 
BRST-invariant at $(l-1)$ loops; 
then the $l$-loop functional generator of the $1PI$ Green functions
renormalized up to $(l-1)$ loops, called $\Gamma^{\e(l)}$,
satisfies the Slavnov-Taylor identities, which in   
the Zinn-Justin form \cite{Zinn} read
\begin{eqnarray}\label{Zinn}
S^{(0)}*\Gamma^{\e(l)} + \Gamma^{\e(l)}*S^{(0)} =
-\sum_{m=1}^{l-1} \Gamma^{\e(m)} * \Gamma^{\e(l-m)}
\end{eqnarray}
In MS-SNDR $\Gamma^{\e(l')}$, for $l' < l$, is finite and belongs to
$T_0$ for $\e = 0$;
 due to (\ref{T1}), for $\e = 0$
the right-hand-side of (\ref{Zinn})
belongs also to $T_0$; it follows that the $T_1$ local term of 
$\Gamma^{\e(l)}$ is BRST invariant, as well as its poles;
therefore the bare action is BRST invariant also at $l$ loops.

\subsection{Comparison with other NDR schemes}

The NDR prescriptions proposed in the past have, as regularization schemes,
some inconsistency, so that they require some care in using them,
usually on a diagram-by-diagram basis,
while in a true regularization scheme one should be able to get 
unambiguous results by using the bare action and the Feynman rules,
whatever computational routine is followed.

Furthermore no proof has been given, in any of these schemes, that
the renormalized Green functions are equal, modulo finite local
renormalization terms , to those in some consistent scheme, e.g. the BMHV
scheme, at all orders in perturbation theory.

The issue of preserving the Ward identities comes only after settling
these points, and it is solved by adding local finite terms to the
bare action. In absence of a consistent regularization system
 it is not true that checking the Ward identities on the relevant vertices
one is guaranteed that the theory is renormalized consistently.
In fact consider a set of `renormalized' Green functions which differs
from any consistent set of renormalized Green functions by  non-local
quantities \footnote{
i.e. such that there is no renormalization group transformation
transforming one set of Green functions in the other; one can
construct easily similar pathological cases by choosing, within a
consistent regularization system, two different one-loop 
sets of renormalization conditions, say $A$ and $B$, and by computing 
a subset of the two-loop Green functions in system $A$, another subset
in system $B$; the resulting mixing of Green functions differs by
non-local terms from the set of Green functions computed within a single
scheme. Working without a consistent regularization system without
great care it is likely that a similarly inconsistent system 
is produced at some point.}
Then check the Ward identities on the relevant terms defined
at a subtraction point. If there are no anomalies, one can add local
finite counterterms to restore the Ward identities on the relevant
terms. Clearly the new set of Green functions is still inconsistent,
since non-local $l$-loop terms cannot be canceled by local $l$-loop 
counterterms.

The dangers of using algebraically inconsistent regularization systems
have been reviewed in \cite{Bonn}, where it is emphasized 
the fact that the renormalization conditions are compatible with the 
Ward identities does not guarantee the validity of the quantum action 
principle, in absence of a true regularization scheme. 

Let us consider briefly some of the NDR schemes discussed in the literature.

To avoid confusion with
previously defined objects, call $\Gamma^5$ the algebraically undefined
$\g^5$-like object; $\g^5$ is defined in (\ref{g5}).

In \cite{CFH} $\Gamma^5$ is chosen to be anticommuting in open fermionic
lines and in loops with an even number of $\Gamma^5$, while in the
loops with an odd number of $\Gamma^5$ the trace rules are given modulo 
some evanescent term, to be fixed by the Ward identities.
The authors emphasize that this is not a true regularization scheme,
but that it is a convenient prescription for one-loop computations.

In \cite{AZ} $\Gamma^5$ is chosen to be anticommuting in open fermionic
lines and in loops with an even number of $\Gamma^5$, while in the
loops with an odd number of $\Gamma^5$ one reduces the traces with an odd
number of fermions, before evaluating them,
 to the case of one $\Gamma^5$ only, by using anticommuting 
$\Gamma^5$;
this rule does not preserve the Bose symmetry.

In \cite{Town} a non-cyclic definition of the trace is used, together
with a totally anticommuting $\Gamma^5$. Bose symmetry is broken in this
scheme; to avoid this problem one can add the rule of choosing an
appropriate reading point in the graphs. 

\begin{figure}
\centering
\includegraphics[width=4cm]{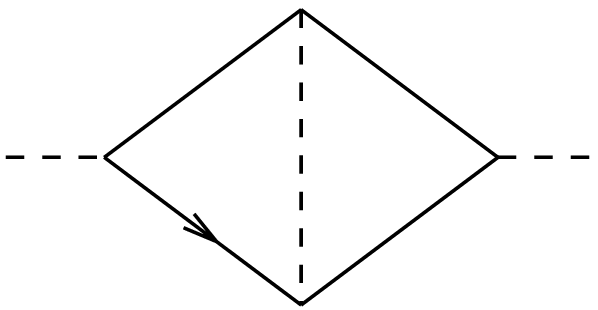}
\caption{}
\label{fig:3}
\end{figure}

In \cite{Veltman} it is suggested the use of $\g^5$ together with 
a trace rule which, in the case of
even number of $\g^5$, acts as if the $\g^5$ were naive, while in the
case of odd number of $\g^5$ it acts as the BMHV trace. In this case
there are ambiguities in the treatment of objects like
$tr (\{\g^5,\g_\m \} \{\g^5,\g_\m \}) \g_\n \g_\r$, which is zero or not
depending on the fact that $(\{\g^5,\g_\m \} \{\g^5,\g_\m \})$ is 
given or not its value $4 \e$. 
For instance using this trace rule
the renormalization of the two-loop graph
in Figure \ref{fig:3}  requires some care due to this ambiguity.

Trueman \cite{Trueman} has used the trick of renormalizing the axial
vector-fermion-fermion vertex as the corresponding 
vector-fermion-fermion vertex multiplied by $\g^5$, as a way of
restoring the Ward identities and for treating $\g^5$ as if it were
anticommuting, while keeping the BMHV rules.
Unlike the other prescriptions reviewed in this subsection,
this one is consistent; however,
apart from the fact that this trick has been defined only for certain
cases, it has the drawback of using the BMHV scheme, with all its
spurious anomalies (i.e. finite counterterms which must be added to
restore the Ward identities); for instance in the case of the Yukawa theory
in \cite{PRR} a trick similar to the one proposed by Trueman has been
used, with the result that the number of spurious anomalies is
conspicuous.
Trueman's trick works well as long as one is able to make computations
without resorting to the full form of the bare action, implicitly
defined; for a three-loop application of this trick see \cite{Larin}.

Let us finally describe the MS-SNDR as a NDR
prescription. 

$\Gamma^5$ is chosen to be anticommuting in open fermionic
lines and in loops with an even number of $\Gamma^5$, while in the
loops with an odd number of $\Gamma^5$ it is considered to be equal to
 $\g^5$.

Apart from subtracting the poles, the MS-SNDR requires a finite 
subtraction  in the fermionic subdiagrams which occur in the fermionic
traces with an odd number of $\Gamma^5$; this finite subtraction
is chosen such that the corresponding fermionic renormalization part 
has the same value, apart from evanescent terms, as the
fermionic renormalization part occurring in an open fermionic line, or
in a closed fermionic line with an even number of $\Gamma^5$
(minimal subtraction of the $T_1$ terms).

In MS-SNDR the BMHV $\g^5$ does not appear in the tree-level action,
but can be produced in divergent graphs, by application of the
Levi-Civita tensor on the gamma matrices. The resulting counterterms
must not be confused with those with $\Gamma^5$; in particular the
counterterms $(\Gamma^5 - \g^5)/\e$ cannot be neglected
($(\eta- \g^5)/\e$ is not weakly evanescent).

\vskip 1cm

I thank M. Raciti and F. Riva for help and discussions.

\end{document}